\documentclass[12pt]{iopart}
\usepackage{iopams}
\usepackage{graphicx}
\usepackage{color}
\usepackage{amsfonts}
\input amssym.def 
\input amssym

\begin{document}


\title[]{Geometric contextuality from the Maclachlan-Martin Kleinian groups}

\author{ Michel Planat
}

\vspace*{.1cm}
\address
{
 Institut FEMTO-ST, CNRS, 15 B Avenue des Montboucons, F-25033 Besan\c con, France. ({\tt michel.planat@femto-st.fr})}

\vspace*{.2cm}

\vspace*{.1cm}
\normalsize

\begin{abstract}

There are contextual sets of multiple qubits whose commutation is parametrized thanks to the coset geometry $\mathcal{G}$ of a subgroup $H$ of the two-generator free group $G=\left\langle x,y\right\rangle$. One defines geometric contextuality from the discrepancy between the commutativity of cosets on $\mathcal{G}$ and that of quantum observables.
It is shown in this paper that Kleinian subgroups $K=\left\langle f,g\right\rangle$ that are non-compact, arithmetic, and generated by two elliptic isometries $f$ and $g$ (the Martin-Maclachlan classification), are appropriate contextuality filters. Standard contextual geometries such as some thin generalized polygons (starting with Mermin's $3 \times 3$ grid) belong to this frame. The Bianchi groups $PSL(2,O_d)$, $d \in \{1,3\}$ defined over the imaginary quadratic field $O_d=\mathbb{Q}(\sqrt{-d})$ play a special role.


 

\end{abstract}


\section{Introduction}
\noindent

In a previous work, a concept of \lq geometric contextuality' fitting that of \lq quantum contextuality' of a set of observables has been identified for the first time \cite{Planat2015}.  Impossible assignments -a la Kochen-Specker- of eigenvalues $\nu$ for observables $A$ happen in a finite geometry $\mathcal{G}$ whose points are the observables and lines are commuting sets of them. For two and three qubits, Mermin's $3 \times 3$ grid and pentagram are the smallest such contextual configurations \cite{Planat2013} which are also present in the quantum pigeonhole effect \cite{YuandOh2015}. In quantum contextuality, a geometry is parametrized by the observables whilst in geometric contextuality the points of $\mathcal{G}$ are representatives of coset classes of a subgroup $H$ of the free group $G=\left\langle x,y \right\rangle$ on two generators. The geometry $\mathcal{G}$ itself is stabilized under the action of cosets - the latter action is known as a Grothendieck's \lq dessin d'enfant' \cite{Dessins2014}. Geometric contextuality arises when it becomes impossible to find all lines of $\mathcal{G}$ with commuting cosets, i.e. not all lines have their points/cosets satisfying the coset commutation law $[x_1,x_2,\cdots, x_p]=e$ whatever the ordering of the $p$ points.

\subsubsection*{Kleinian groups of the MM-census}

 With this new coset approach in hands, what basic features should possess a subgroup $H$ of $G$ to generate a contextual $\mathcal{G}$? In this paper, we show that the generic \lq contextual subgroups' are Kleinian, non-compact, arithmetic and arise from two elliptic generators. The work of C. Maclachlan and G.~J. Martin establishes that there are $21$ such Kleinian groups $K$ (herafter denoted the MM-census) \cite{Maclachlan2001}. We examine the small index subgroups of $K$ in a few classes of the MM-census and how they help to clarify our topic within the new frame of $3$-orbifolds $\mathbb{H}^3/K$. 

A Kleinian group $K$ is a discrete subgroup of $PSL(2,\mathbb{C})$, the group of all orientation-preserving isometries of the $3$-dimensional space $\mathbb{H}^3$. A non-compact group $K$ is such that the corresponding orbifold $\mathbb{H}^3/K$ is non-compact. A Kleinian group $K$ is called arithmetic if it is commensurable with the group of units of an order of quaternion algebra $A$ ramified at all real places over a number field k with exactly one complex place. Arithmetic Kleinian groups have finite covolume \cite[Sec. 8.2]{MaclachlanReid}. There are finitely many two-generator Kleinian groups $K=\left\langle f,g\right\rangle$ that are non-compact and arithmetic provided the isometries $f$ and $g$ are elliptic, with degrees $p$ and $q$ respectively, satisfying $\mbox{tr}^2(f)=4 \cos^2 (\pi/p)$, $\mbox{tr}^2(g)=4 \cos^2 (\pi/q)$  with parameter $\gamma=\mbox{tr}[f,g]-2$. 

It is known that a quadratic imaginary number field $kK=\mathbb{Q}(\sqrt{-d}$ and a quaternion algebra $AK\equiv M_2(\mathbb{Q}(\sqrt{-d}))$ are associated to a non-compact arithmetic Kleinian group $K$. It is also known that $K$ is commensurable, up to conjugacy, with the Bianchi group $PSL(2,O_d)$, where $O_d$ is the ring of integers in $\mathbb{Q}(\sqrt{-d})$. For groups $K$ of the $MM$-census, $d \in \{1,3\}$ \cite{Maclachlan2001}.

Table 1 summarizes the Kleinian groups that serve as models of geometric contextuality as described in the subsequent sections 1 to 5.

\begin{table}[ht]
\begin{center}
\caption{Parameters of a few classes of Kleinian groups in the MM-census. The group $K=K^{(i)}$ means the Kleinian group in class $i$ of the MM-census, $p$ and $q$ are degrees of the elliptic isometries $f$ and $g$ generating $K$, $-d$ is the discriminant of the Bianchi group  to which $K$ is commensurable, $\gamma=\mbox{tr}[f,g]-2$, the notation $\mathcal{G}$ means a geometry occurring in the relevant finite index subgroup of $K$ and Sec. is the section of our paper where $K$ occurs.}
\begin{tabular}{||l|rcrcl|lr|}
\hline \hline
group& $p$ & $q$ & $d$  & $\gamma$ & covolume & $\mathcal{G}$  & Sec.\\
\hline\hline
$K^{(4)}$ & $2$ &  $4$&  $1$ & $-1+i$ & $0.45798$  & Hesse, Petersen & 2\\
$K^{(19)}$ & $4$ &  $6$&  $3$ & $-1$ & $0.21145$  & $GQ(2,1)$ (Mermin square) & 3\\
$K^{(1)}$ & $2$ &  $3$&  $3$ & $(-3+\sqrt{3}i)/2)$ & $0.33831$  & Mermin pentagram, $GH(2,1)$ & 4\\
$K^{(5)}$ & $2$ &  $4$&  $1$ & $-2+i$ & $0.91596$  & $GO(2,1)$, \lq\lq$GO(2,4)$"& 5\\
$K^{(2)}$ & $2$ &  $3$&  $3$ & $(-1+\sqrt{3}i)/2$ & $0.67664$  & $GH(2,1)$, \lq\lq $J_2$" & 5\\
\hline
\hline
\end{tabular}
\end{center}
\end{table}

\subsubsection*{Generalized polygons}

It is noticeable that many of the generic contextual configurations filtered in the MM-census are \lq thin' generalized polygons.
A generalized polygon (or generalized $n$-gon) is an incidence structure between a discrete set of points and lines whose incidence graph has diameter $n$ and girth $2n$ \cite{psm}. The definition implies that a generalized $n$-gon cannot contain $i$-gons for $2 \le i <  n$ but can contain ordinary $n$-gons. A generalized polygon of order $(s,t)$ is such that every line contains $s+1$ points and every point lies on $t+1$ lines. A projective plane of order $n$ is a generalized $3$-gon. The generalized $4$-gons are the generalized quadrangles. Generalized $5$-gons, $6$-gons, etc are also called generalized pentagons, hexagons, octagons, etc. According to Feit-Higman theorem, finite generalized $n$-gons with $s>1$, $t>1$ may exist only for $n \in \{2,3,4,6,8\}$ \cite{Tits2002}. Such structures are relevant for quantum contextuality as shown in \cite{Planat2015}. A \lq thin' generalized polygon is such that $s=1$ ot $t=1$. Below we will meet the thin generalized quadrangle $GQ(2,1)$ (i.e. the $3 \times 3$ grid, also named Mermin's square), the  hexagon $GH(2,1)$ and the octagon $GO(2,1)$. These thin polygons have three points on their lines and valency two for their points.

\section{Bell's quadrangle, the Hesse configuration and the Petersen graph}
\noindent

Even the unprejudiced reader may find our language esoteric before we settle that the smallest non trivial $\mathcal{G}$ is the ordinary quadrangle (or square graph) and that it is relevant to Bell's theorem, a basic component of quantum contextuality, as was repetitively justified in our recent papers \cite[Sec. 3]{Dessins2014},\cite[Sec. 2]{Moonshine2015}.

A finite representation of the Kleinian group leading to the {\it dihedral} group $D_4$ of the quadrangle may be taken as that of class $4$ (the same results for the class $6$) in the MM-census (from now the identity element $e$ is denoted $1$)
$$K^{(4)}=\left\langle x,y|  y^2=x ^4=[(yx^{-1})^2(y^{-1}x)^2]^2=1\right\rangle.$$

\begin{figure}[ht]
\centering 
\includegraphics[width=4cm]{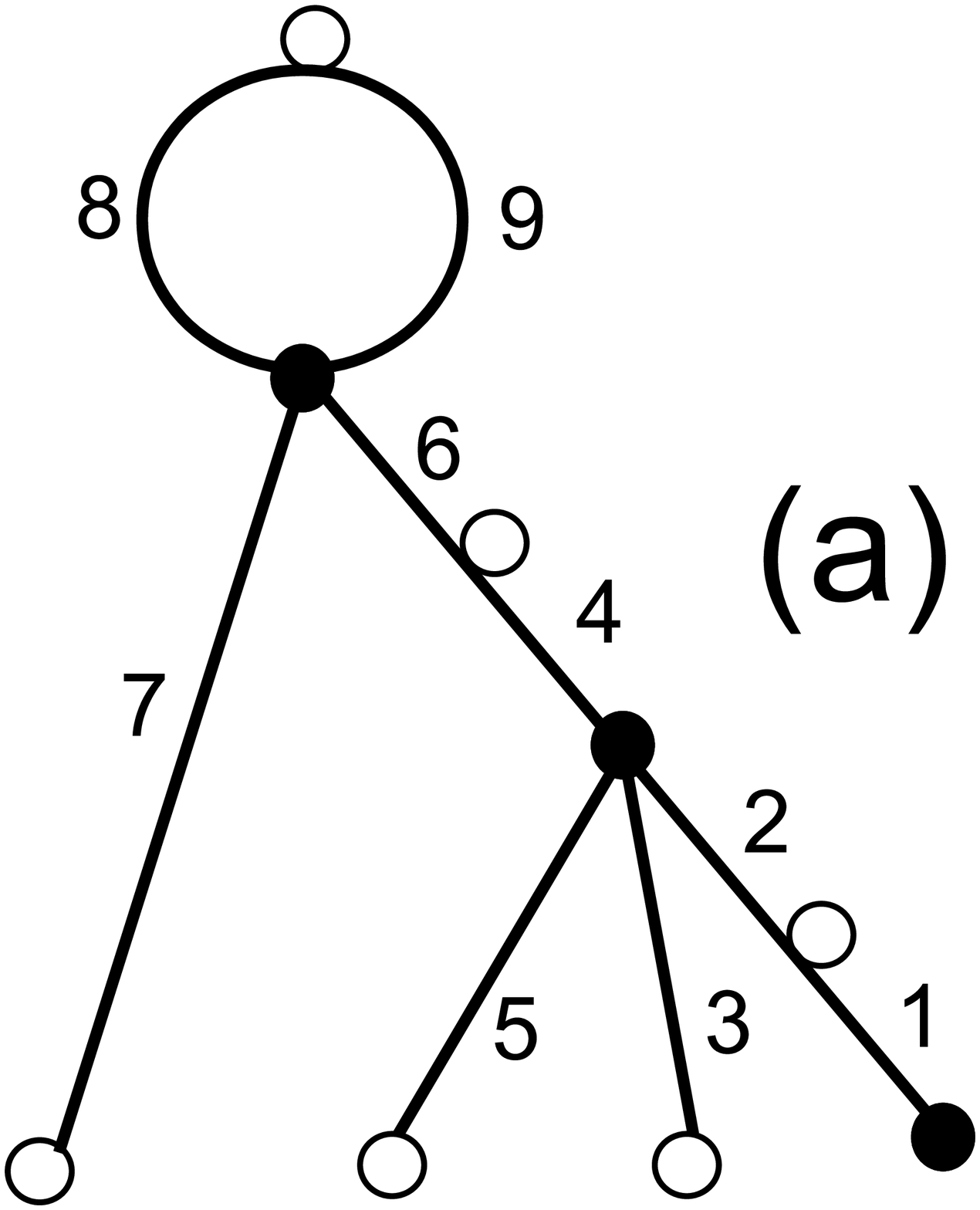}
\includegraphics[width=4cm]{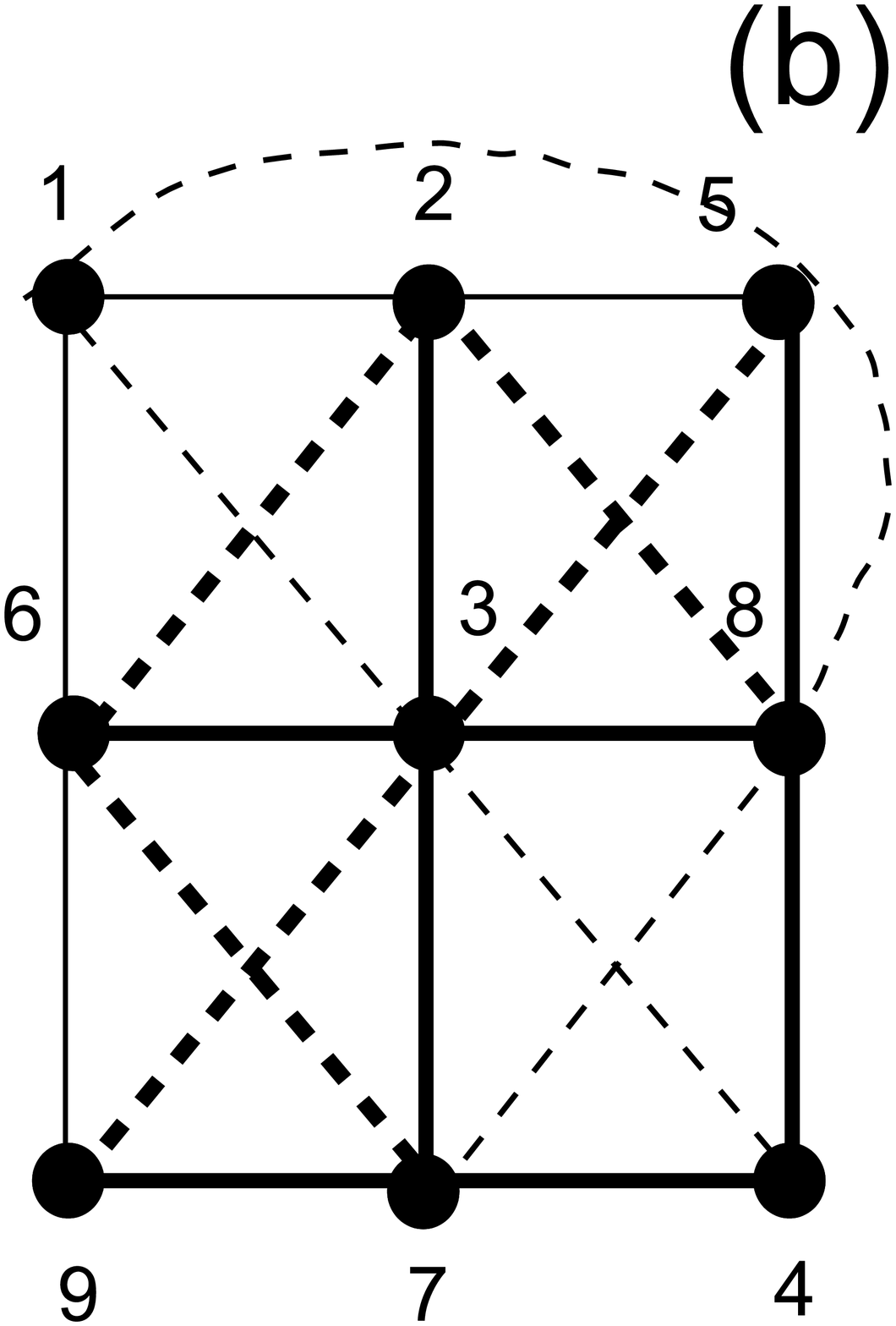}
\caption{ The dessin d'enfant (a) arising from the first subgroup of index $9$ of the Kleinan group $K^{(4)}$ and the Hesse configuration (b) that it stabilizes. The configuration consists of the union of a $3 \times 3$ grid (plain lines) and another one that is skewed to it (dotted lines). Non-straight lines are not all all drawn. The coordinatization is in terms of coset representatives of the subgroup $K^{(4)}$ in the free group $G$ as follows: $1=e,~ 2=y,~ 3=yx,~ 4=yx^{-1},~ 5=yx^2,~ 6=yx^{-1}y,~ 7=yx^{-1}yx,~ 8=(yx^{-1})^2,~ 9=yx^{-1}yx^2$.  Only the thin lines (passing through $e$) have their cosets mutually commuting.
 }
\end{figure}

There are just four subgroups of index $4$ of $K^{(4)}$, of permutation representation $P_1=\left\langle (2,3),(1,2)(3,4)  \right\rangle$, $P_2=\left\langle (1,2)(3,4), (2,3) \right\rangle$, $P_3=\left\langle (1,2,4,3),(1,2)(3,4) \right\rangle$ and $P_4=\left\langle  (1,2,4,3),(2,3)\right\rangle$. These groups are the ones calculated in \cite[Sec. 3]{Dessins2014} (where all details: the signature, the cycle structure and the Belyi function of the corresponding dessin d'enfant are made explicit). The MM-filtering is ineffective here since the free group $G$ only leads to these four cases. Denoting $\mathbb{H}^3$ the upper-half space, what we learn is that the physics of Bell's theorem relates to the orbifold $\mathbb{H}^3/K^{(4)}$ whose graph contains four crossings, see \cite[Fig. 3]{Maclachlan2001}.

There are two subgroups of index $9$ of $K^{(4)}$ that stabilize the Hesse configuration, already found to be important in the context of the Kochen-Specker theorem \cite{Bengtsson2012}. The permutation representation of the first case is illustrated as the dessin d'enfant in Fig. 1a, that is $P=\left\langle (2,3,5,4)(6,7,8,9), (1,2)(4,6)(8,9)\right\rangle$ of order $144$. The attached Hesse configuration is shown in Fig. 1b. Following the definition in our paper \cite{Planat2015}, it is maximally contextual since all lines except the ones passing through the neutral element have non-commuting cosets. Incidentally, the Hesse configuration consists of two Mermin squares skewed to each other, see also \cite[Sec. 4.4 and Fig. 8]{Dessins2014}.

\begin{figure}[ht]
\centering 
\includegraphics[width=5cm]{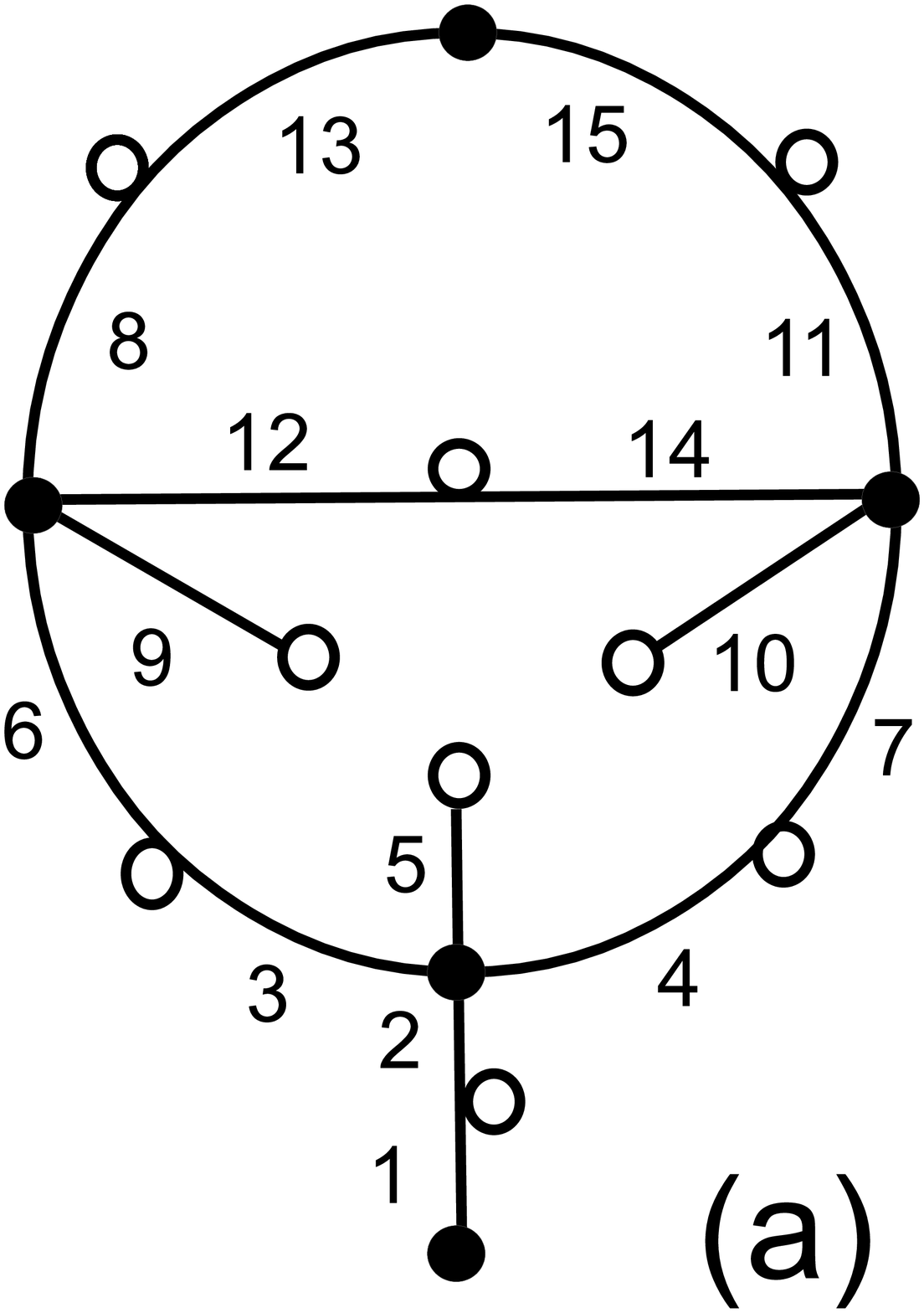}
\includegraphics[width=5cm]{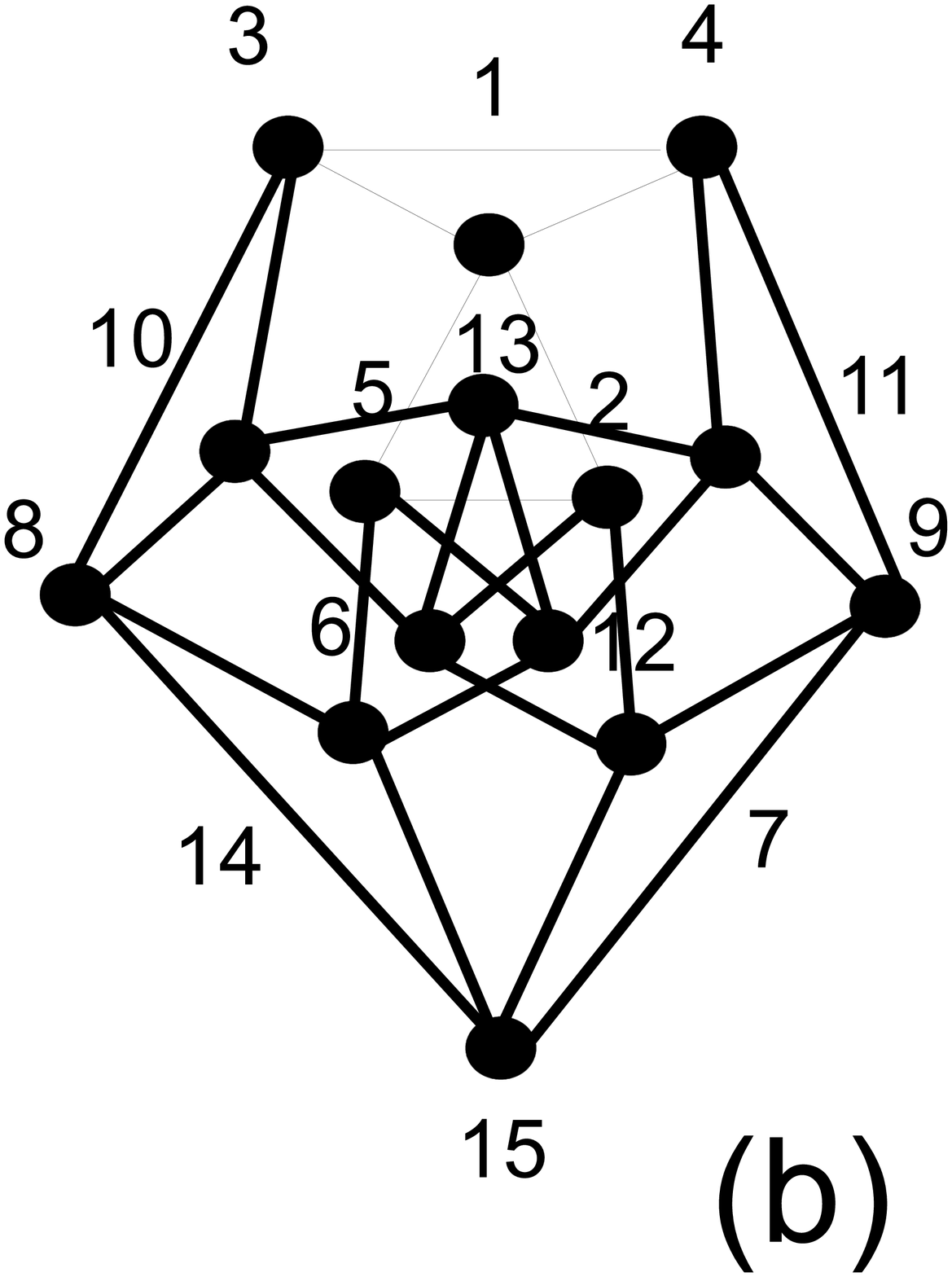}
\caption{ The dessin d'enfant (a) arising from the first subgroup of index $15$ of the Kleinan group $K^{(4)}$ and the line graph of the Petersen graph (b) that it stabilizes. The list of coset representatives  corresponding to the labelling $\{1,2,\cdots,15\}$ is as follows $[e,y,yx,yx^{-1},yx^2,yxy,yx^{-1}y,(yx)^2,yxyx^{-1},yx^{-1}yx,(yx^{-1})^2,yxyx^2,yxyx^{-1}y,
\newline yx^{-1}yx^2,yx^{-1}yxy]$. Only the thin triangles (passing through $e$) have their cosets mutually commuting.
 }
\end{figure}

There are two subgroups of index $10$ of $K^{(4)}$ that are isomorphic to the symmetric group $S_5$. They stabilize the Petersen graph and simultaneously the Desargues configuration -the latter geometry is found to be maximally contextual- see also \cite[Fig. 10 and Fig. 11]{Dessins2014} for other generating dessins.
Here we focus on the two subgroups of index $15$ of $K^{(4)}$ that stabilize the line graph of the Petersen graph. The permutation representation for the first case is encoded in the dessin of Fig. 2a and the corresponding graph is on Fig. 2b. The coset labeling of the graph is maximally contextual as before for the Hesse labeling.

\section{Mermin's square}
\noindent

As announced in the opening section, Mermin's $3 \times 3$ grid is a critical prototype of quantum as well as geometic contextuality \cite[Fig. 3a]{Planat2015}.
Class 19 of the MM-census is the one used to recover this geometry (class 20 is an alternative class). The finite representation of the corresponding Kleinian group is as follows (with the misprint in \cite[Tab. 1]{Maclachlan2001} corrected)
$$K^{(19)}=\left\langle x,y|  y^4=x ^6=[x,y]^3=([y,x]*y)^2 =(y^{-1}*[y,x])^2=   (x^{-1}*[y,x]*y)^2=1\right\rangle.$$
The dessin leading to the $3 \times 3$ grid is shown in Fig. 3a and the resulting grids are in Fig. 3b (non-contextual) and Fig. 3c (contextual).

\begin{figure}[ht]
\centering 
\includegraphics[width=5cm]{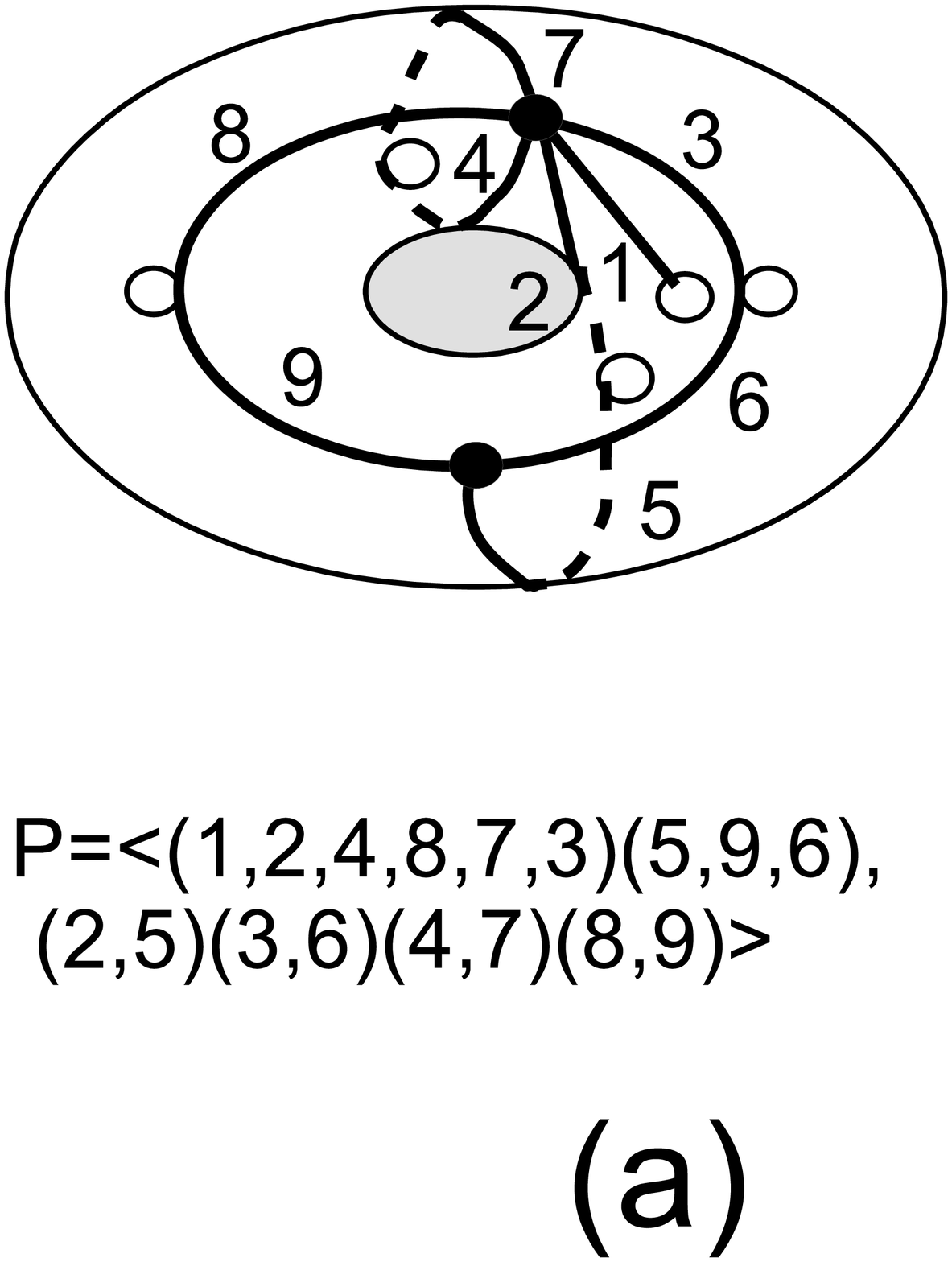}
\includegraphics[width=5cm]{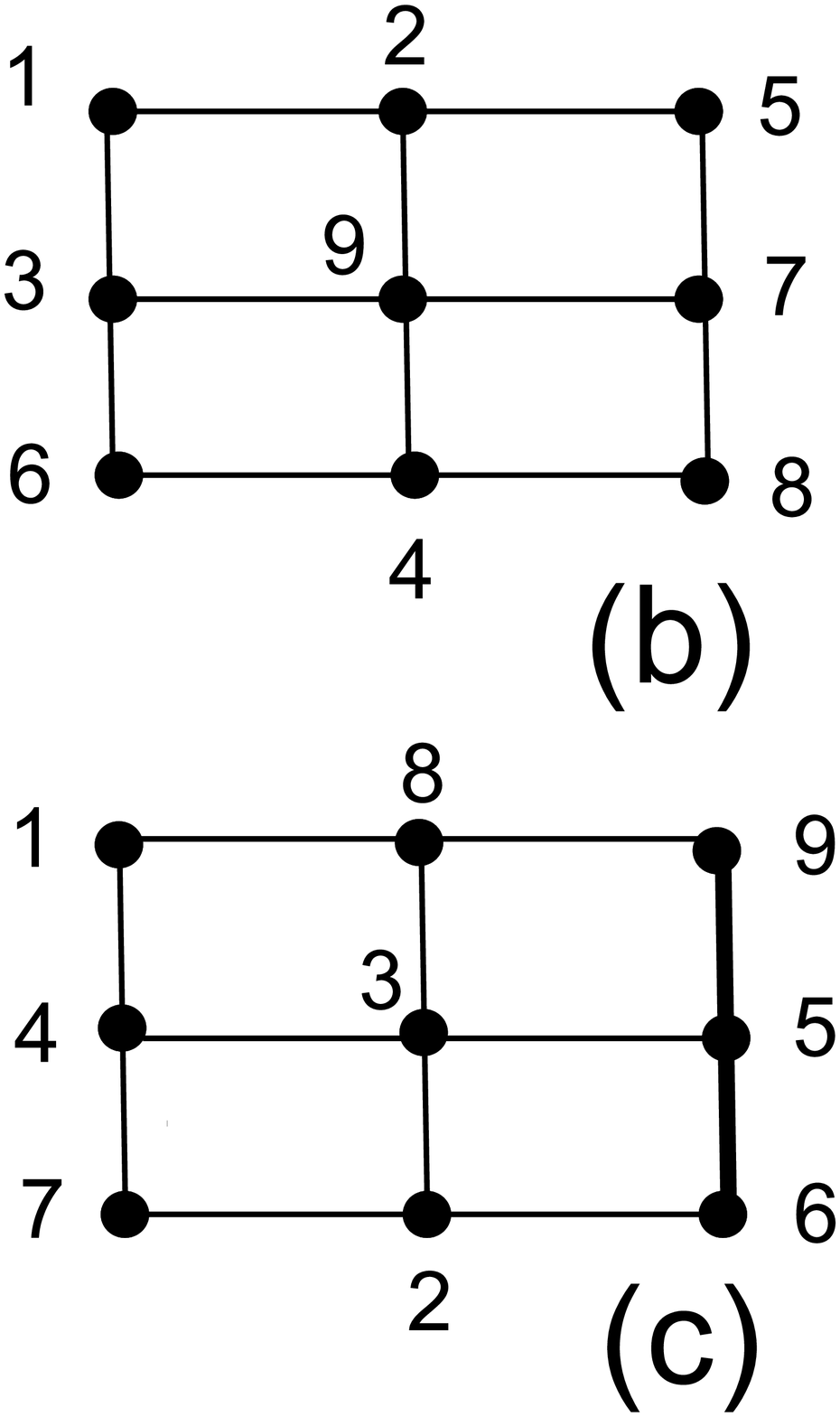}
\caption{The dessin d'enfant (a) arising from the subgroup of index $9$ and permutation group $\mathbb{Z}_3^2 \rtimes \mathbb{Z}_2^2$ in the Kleinian group $K^{(19)}$. This dessin stabilizes the non-contextual grid (b) from the stabilizer subgroup $\mathbb{Z}_1$ and the contextual one (c) from the stabilizer subgroup $\mathbb{Z}_2$. The list of coset representatives  corresponding to the labeling $[1,2,\cdots,9]$ is as follows $[e,x,x^{-1},x^2,xy,x^{-1}y,x^{-2},x^3,xyx]$. The thick line corresponds to non-commuting cosets.
 }
\end{figure}

Subgroups of finite index of the Kleinian group $K^{(9)}$ may also be used to stabilize most multipartite graphs of size larger than $4$ in a contextual way, starting with the octahedron (i. e. the graph $K_{2,2,2}$), $K_{3,3}$, $K_{4,4}$, $K_{2,2,2,2}$( i.e. the $16$-cell), $K_{3,3,3}$ and the highest order ones, except for $K_{5,5}$.

\section{The congruence subgroup $\Gamma(2)$, Mermin's pentagram and the thin generalized hexagon $GH(2,1)$}
\noindent

The first class of the MM-census is modular [i. e. a subgroup of $\Gamma=PSL(2,\mathbb{Z})$] with representation
$$K^{(1)}=\left\langle x,y|  y^2=x ^3=[(yx^{-1})^2(y^{-1}x)^2]^3=1\right\rangle.$$

As for $\mathbb{H}^3/K^{(4)}$, the graph of the orbifold $\mathbb{H}^3/K^{(1)}$ contains four crossings, see \cite[Fig. 3]{Maclachlan2001}. We examine a few noticeable (and contextual) subgroups of $K^{(1)}$.

\subsubsection*{Subgroups of $K^{(1)}$ of index $6$}

There are five subgroups of index $6$ of $K^{(1)}$, they are congruence subgroups of $\Gamma$, one of type $\Gamma(2)$ (see below), one of type $\Gamma_0(5)$ and the other two of level $6$ named \lq$ 6 A^0$' and \lq$ 6 A^1$' in Cummins-Pauli classification \cite{Cummins}.

\begin{figure}[ht]
\centering 
\includegraphics[width=4cm]{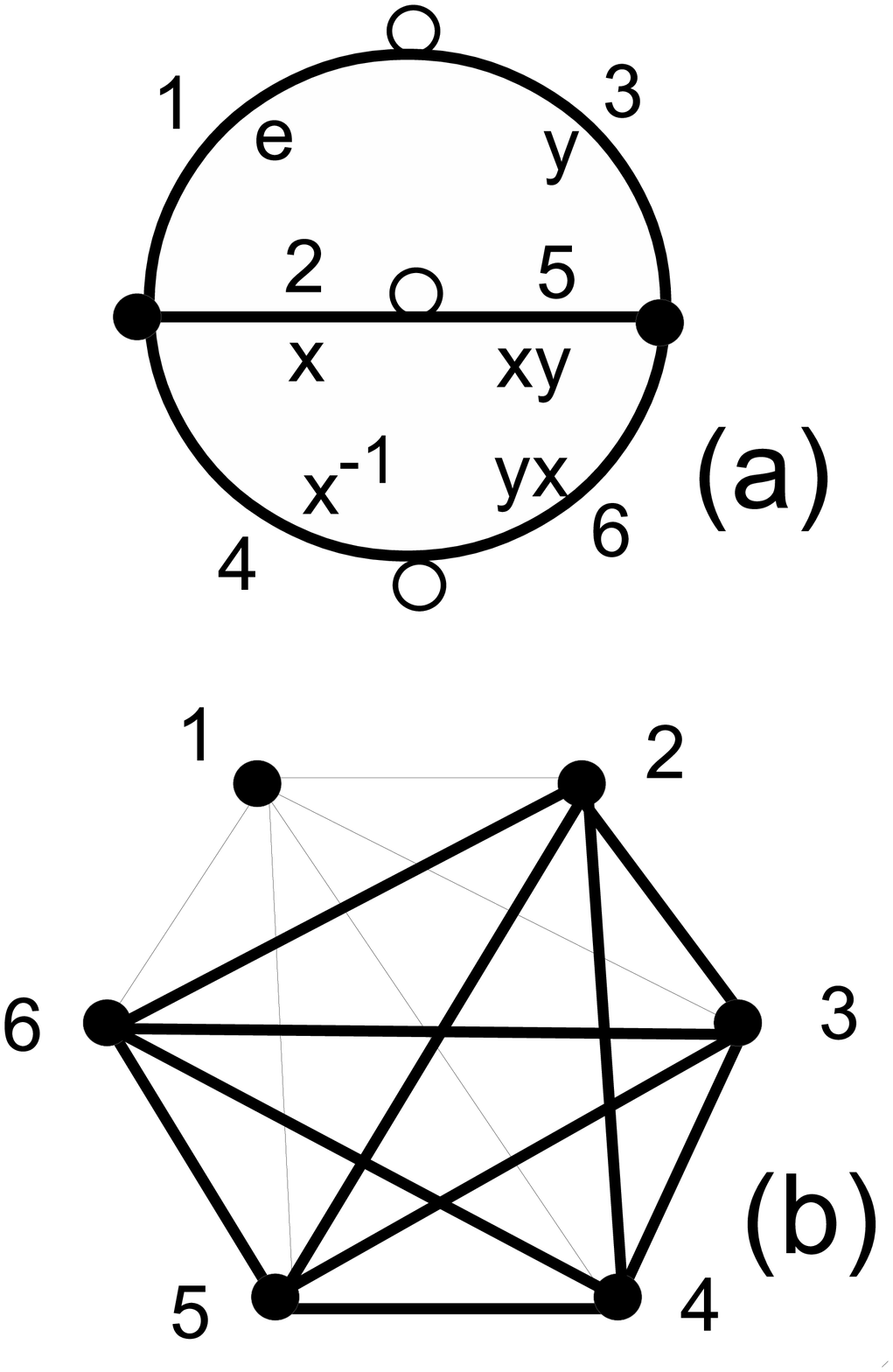}
\includegraphics[width=4cm]{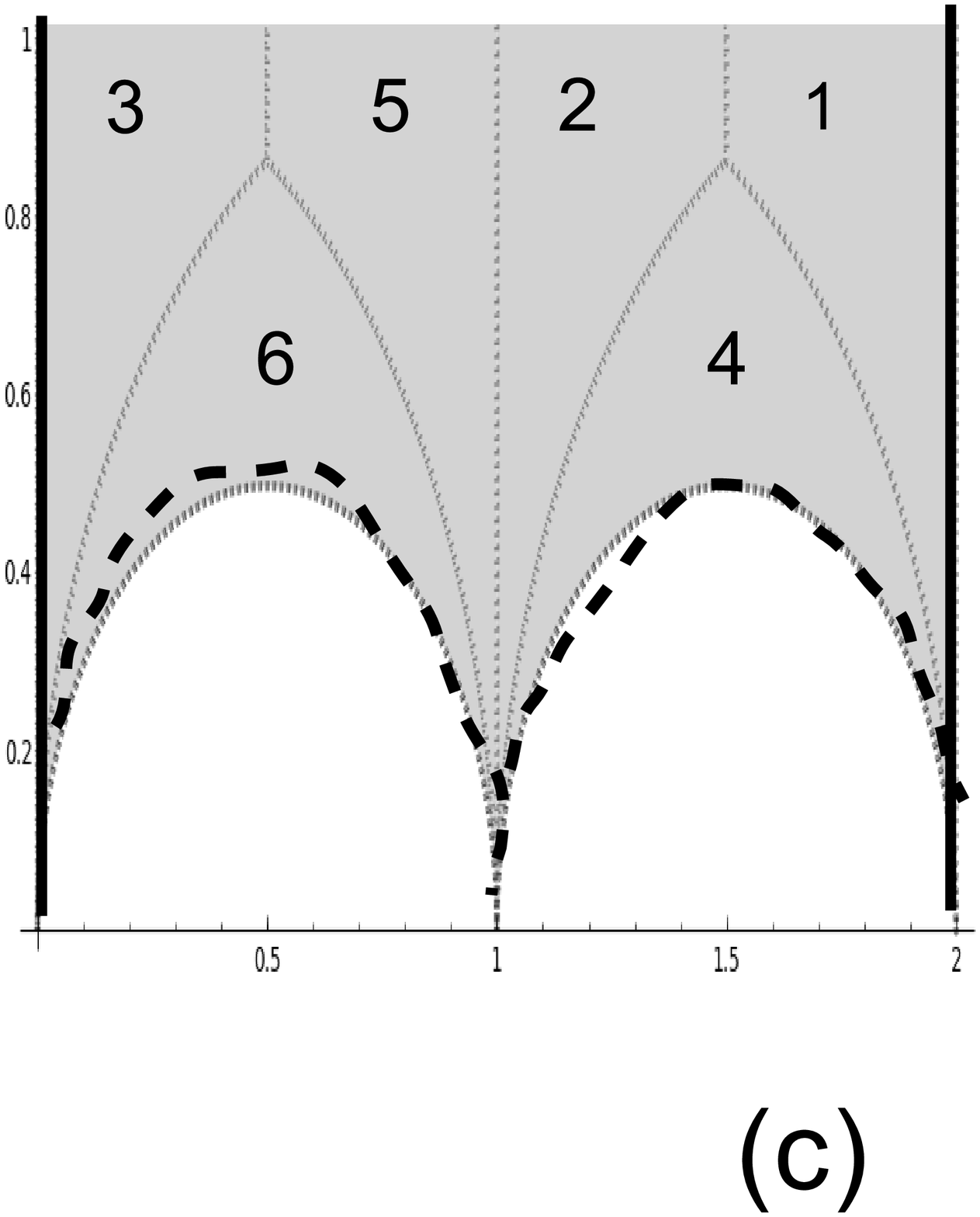}
\caption{The dessin in (a)  arising from the congruence subgroup  of index $6$ of type $\Gamma(2)$ of the Kleinian group $K^{(1)}$. This dessin stabilizes the (maximally contextual) complete graph $K_6$ shown in (b). The thick line corresponds to non-commuting cosets.  The corresponding polygon $\Gamma(2)$ is shown in (c). 
 }
\end{figure}

The dessin $\mathcal{D}$ and the corresponding polygon of type $\Gamma(2)$ are in Fig. 4a and 4c, respectively. The stabilized geometry is a (maximally contextual) complete graph $K_6$ shown in Fig. 4b. The Belyi function for $\mathcal{D}$ is $f(x)=\frac{4}{27} j(x)$, where $j(x)=\frac{(1-x+x^2)^3}{x^2(x-1)^2}$ is the modular invariant \cite[p. 267]{Girondo2012}. 

\subsubsection*{Subgroups of $K^{(1)}$ of index $7$}

There are two subgroups of index $7$ of $K^{(1)}$. They correspond to a permutation group isomorphic to $PSL(2,7)$ (of order $168$) and stabilize a (maximally contextual) Fano plane. The corresponding hyperbolic polygon is of type \lq$7A^0$' in Cummins-Pauli table \cite{Cummins}. The corresponding pictures are not drawn in this paper but the reader can refer to \cite[Fig. 4]{Dessins2014} for details.

\subsubsection*{Mermin's pentagram}

\begin{figure}[ht]
\centering 
\includegraphics[width=6cm]{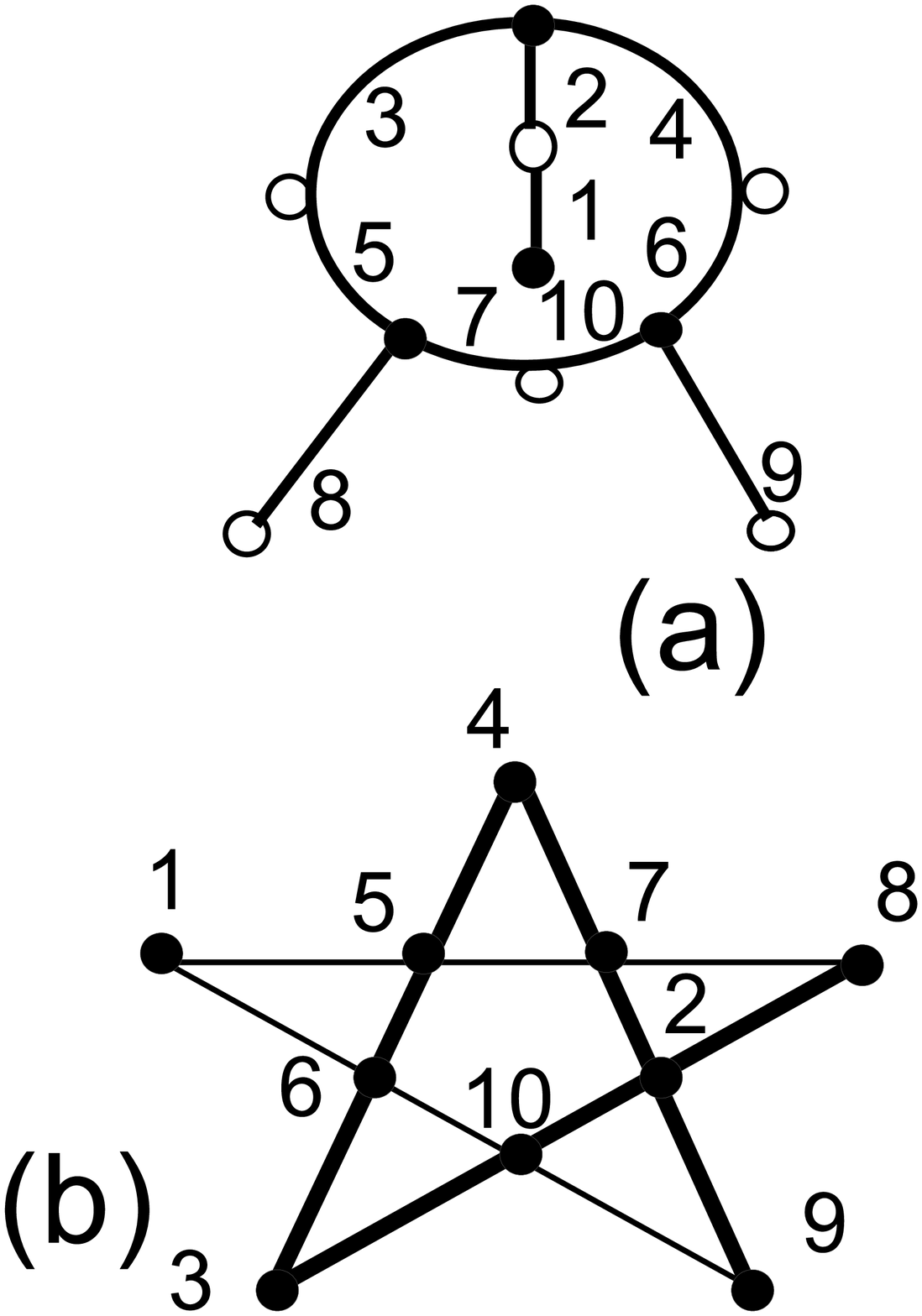}
\includegraphics[width=6cm]{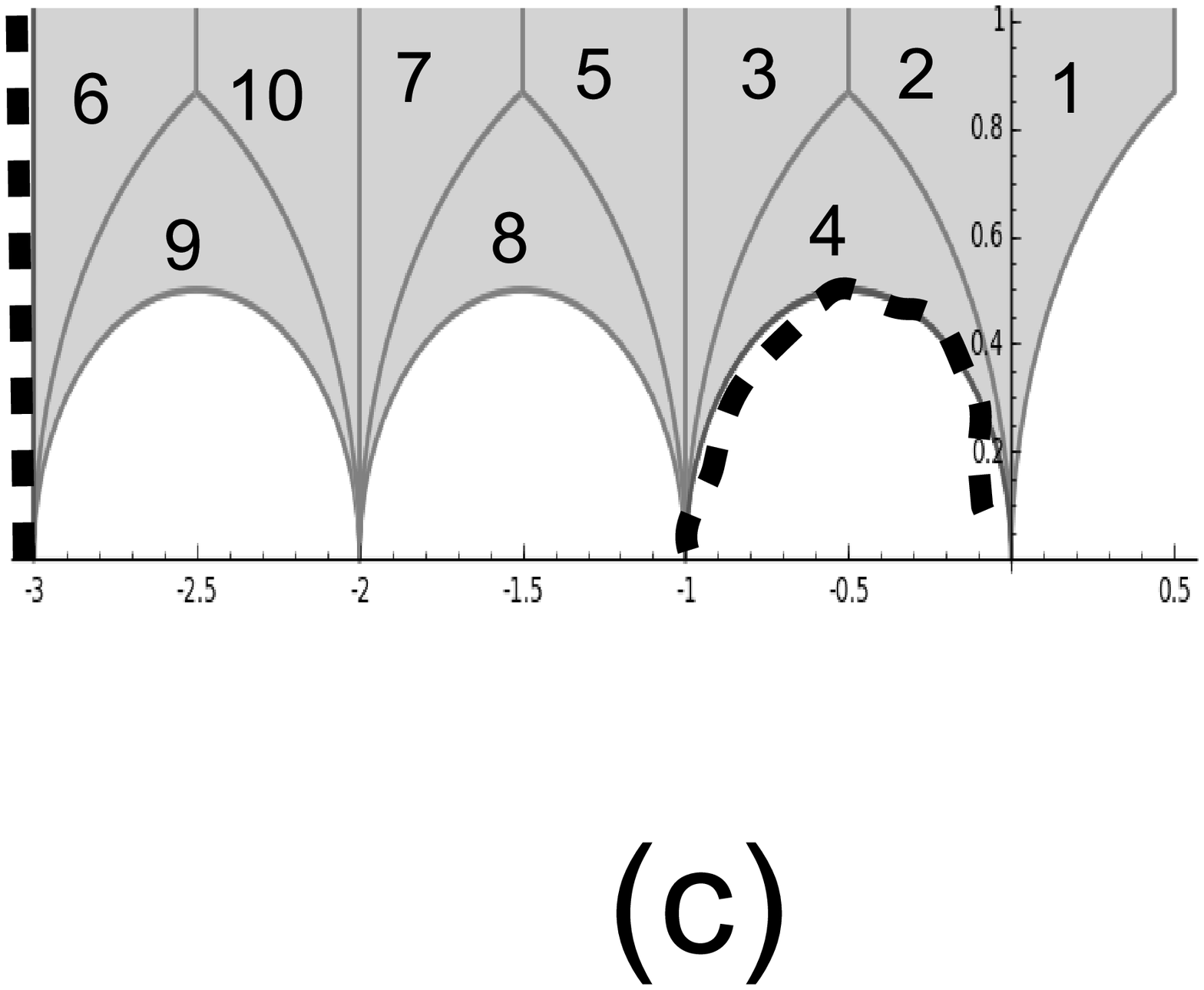}
\caption{The dessin in (a)  arises from the unique subgroup  of index $10$ of the Kleinian group $K^{(1)}$ isomorphic to $S_5$. This dessin stabilizes Mermin's pentagram shown in (b). As it is modular the dessin can also be seen as a tiling of the upper-half plane in (c). The list of coset representatives  corresponding to the labelling $[1,2,\cdots,10]$ is $[e,y,yx,yx^{-1},yxy,yx^{-1}y,(yx)^2,yxyx^{-1},yx^{-1}yx,(yx^{-1})^2]$. The thick lines of (b) correspond to non-commuting cosets. 
 }
\end{figure}

There is just one subgroup of index $10$ of $K^{(1)}$ which is isomorphic to the alternating group $A_5$. The generating dessin, shown in Fig. 5a, stabilizes Mermin's pentagram shown in Fig. 5b, see also \cite[Fig. 3b]{Planat2015}. As $K^{(1)}$ is a subgroup of $\Gamma$, the dessin in Fig. 5a can alternatively be seen as a tiling of the upper-half plane, as shown in Fig. 4c.  The generators $\alpha=
 (2, 3, 4)(5, 7, 8)(6, 9, 10)$ of order three and $\beta=(1, 2)(3, 5)(4, 6)(7, 10)$ of order two of the permutation group $P=\left\langle \alpha,\beta \right\rangle$ build a subgroup $\Gamma'$ of the modular group $\Gamma$ which is a congruence subgroup of level $5$.

It is time to remind how to pass from the topological structure of a modular dessin $\mathcal{D}$ to that of a hyperbolic polygon $\mathcal{P}$ \cite[Sec. 3]{Moonshine2015}. There are $\nu_2$ elliptic points of order two (resp. $\nu_3$ elliptic points of order three) of $\mathcal{P}$, these points are white points (resp. black points) of valency one of $\mathcal{D}$. For the dessin in Fig. 5a, one gets $\nu_2=1$ and $\nu_3=2$. The genus of $\mathcal{P}$ equals that of $\mathcal{D}$, a cusp of $\mathcal{P}$ follows from a face of $\mathcal{D}$, the number $B$ of black (resp. the number $W$ of white) points of $\mathcal{D}$ is given by the relation $B=f+\nu_2-1$ (resp. $W=n+2-2g-B-c$), where $f$ is the number of fractions and $c$ the number of cusps in $\mathcal{P}$. For the dessin in Fig. 5a of cycle structure $[3^31^1,2^41^2,5^2]$ in which the items from left to right are for black points, white points and faces, one has $B=3+1=4$, W=$4+2=6$, $g=0$, $c=2$ and $f=4$.

 The set of cusps for $\Gamma'$ consists of the $\Gamma'$-orbits of $\{\mathbb{Q}\}\cup\{\infty\}$, the cusps are at $-3$ and $\infty$ and they have width $3$.  Here $\Gamma'$ is of type \lq $5C^0$' in Cummins-Pauli classification \cite{Cummins} . We used the software Sage to draw the fundamental domain of $\Gamma'$ thanks to the Farey symbol methodology. Some details about the use of Sage on modular aspects of dessins are given in an essay by Lieven le Bruyn \cite{Lieven2012}. 

\subsubsection*{The thin generalized hexagon $GH(2,1)$}

There is a single subgroup of index $21$ of $K^{(1)}$ isomorphic to $SL(2,7)$. The generating dessin $\mathcal{D}$ is in Fig. 6a, it stabilizes the thin generalized hexagon $GH(2,1)$ with $21$ points and $14$ lines as shown in Fig. 6b. The corresponding hyperbolic polygon $\mathcal{P}$ in $\Gamma$ is shown in Fig. 6c. Following the approach given at the previous subsection, the cycle structure of $\mathcal{D}$ is $[3^7,2^9 1^3,8^24^11^1]$ where $B=7$ and $W=12$ leads to $\nu_2=0$, $\nu_3=3$,  $c=4 $ and $f=8$. In contrast to the previous case, the group $\Gamma'$ attached to $\mathcal{P}$ is not a congruence subgroup of $\Gamma$.

\begin{figure}[ht]
\centering 
\includegraphics[width=6cm]{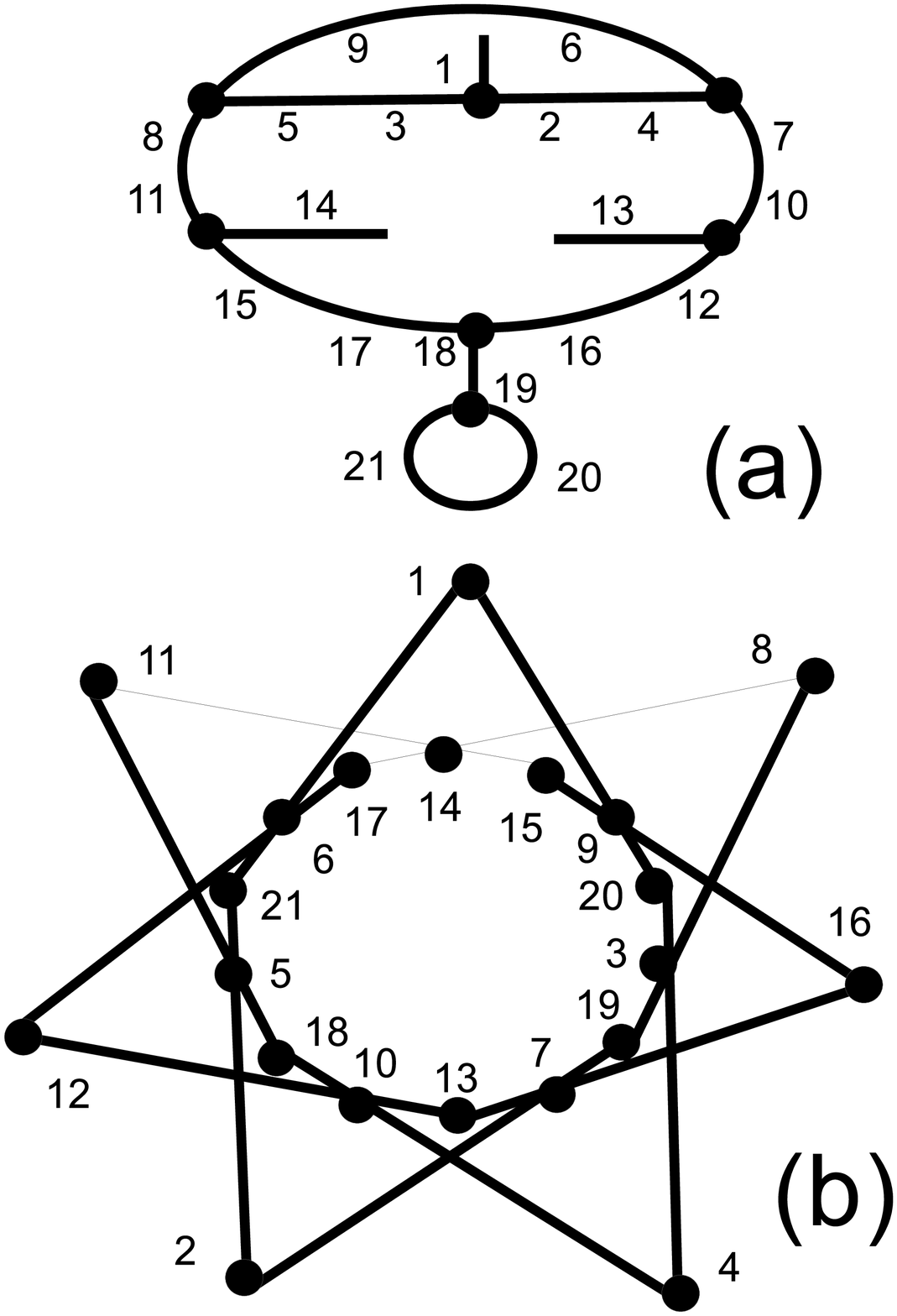}
\includegraphics[width=6cm]{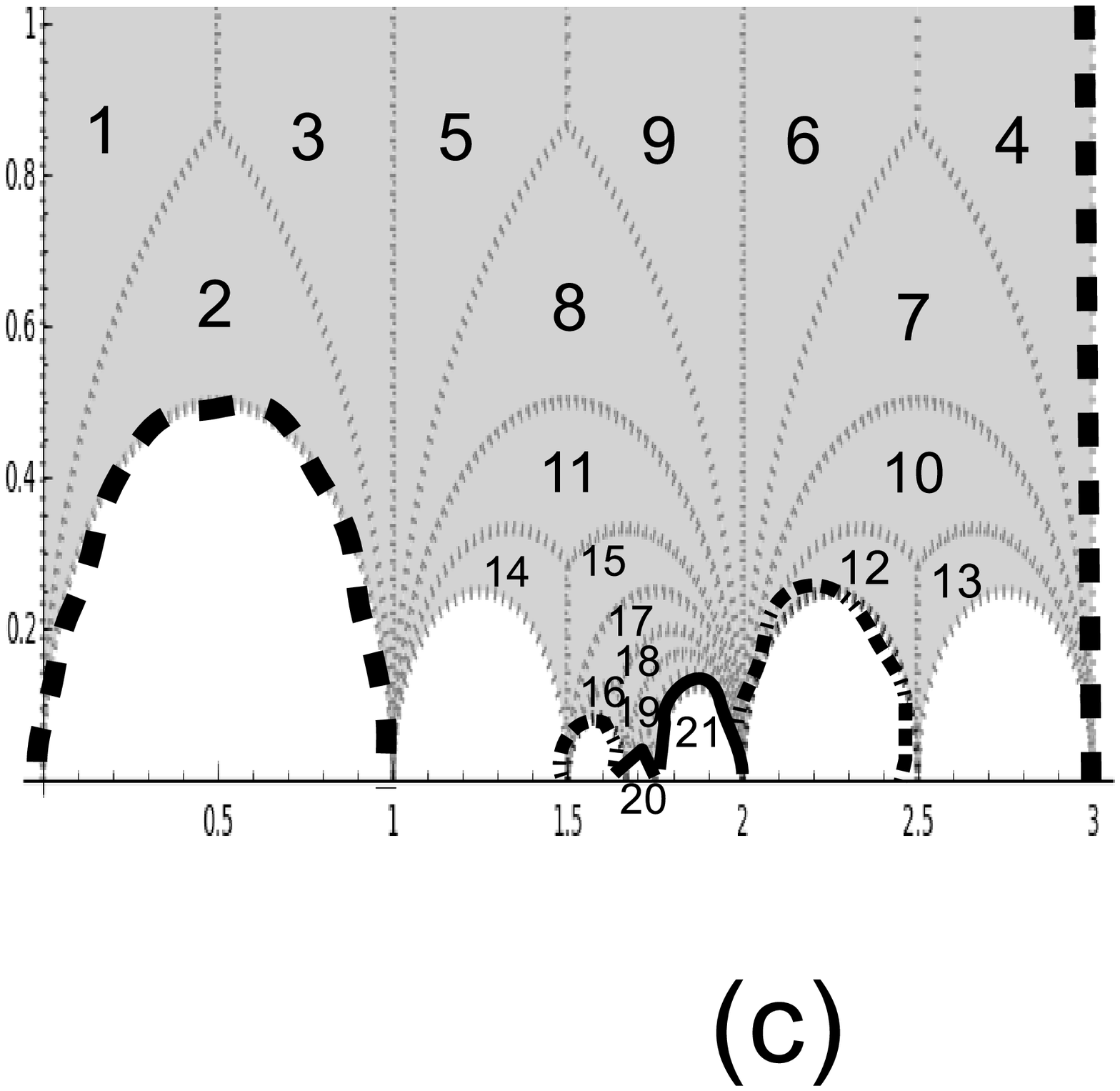}
\caption{The dessin in (a)  arises from the unique subgroup  of index $21$ of the Kleinian group $K^{(1)}$ isomorphic to $SL(2,7)$ (white points are implicit in the drawing). This dessin stabilizes the thin generalized hexagon $GH(2,1)$ shown in (b). As it is modular, the dessin can also be seen as a tiling of the upper-half plane in (c). The list of coset representatives  corresponding to the labelling $1$ to $21$ is $[e, x, x^{-1}, xy, x^{-1}y, xyx, xyx^{-1}, y^x, x^{-1}yx^{-1}, xyx^{-1}y, x^{-1}yxy, xyx^{-1}yx, xyx^{-1}yx^{-1},\newline x^{-1}(yx)^2, x^{-1}yxyx^{-1}, xyx^{-1}yxy,x^{-1}yxyx^{-1}y, xyx^{-1}(yx)^2, xyx^{-1} (yx)^2y, xyx^{-1}(yx)^3 ,\newline xyx^{-1} (yx)^2 y x^{-1} ]$. The thick lines of (b) correspond to non-commuting cosets. 
 }
\end{figure}

\section{The thin generalized octagon $GO(2,1)$ and the Ree-Tits octagon $GO(2,4)$ }
\noindent

The fifth class of the MM-census has the finite representation
$$K^{(5)}=\left\langle x,y| y^2=x ^4=[yxy^{-1}xyx^{-1}]^4=1\right\rangle.$$
The graph of the orbifold $\mathbb{H}^3/K^{(5)}$ contains three crossings \cite[Fig. 3]{Maclachlan2001}.

There is a single subgroup of index $45$ of $K^{(5)}$ isomorphic to $A_6$. The generating dessin in Fig. 7a stabilizes the thin generalized octagon $GO(2,1)$ shown in Fig. 7b.  The latter contains $45$ points and $30$ lines and it is maximally contextual.

\begin{figure}[ht]
\centering 
\includegraphics[width=6cm]{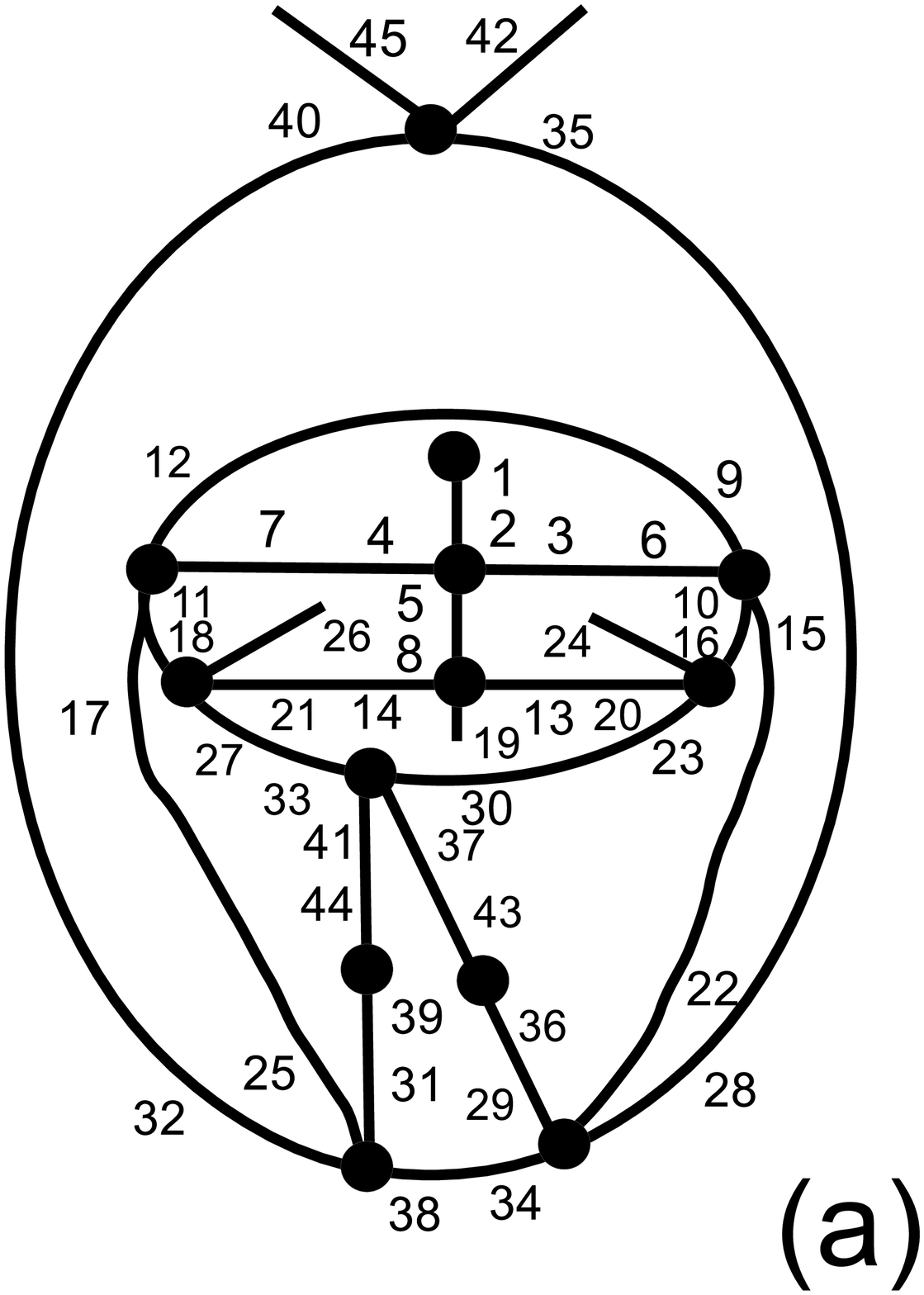}
\includegraphics[width=6cm]{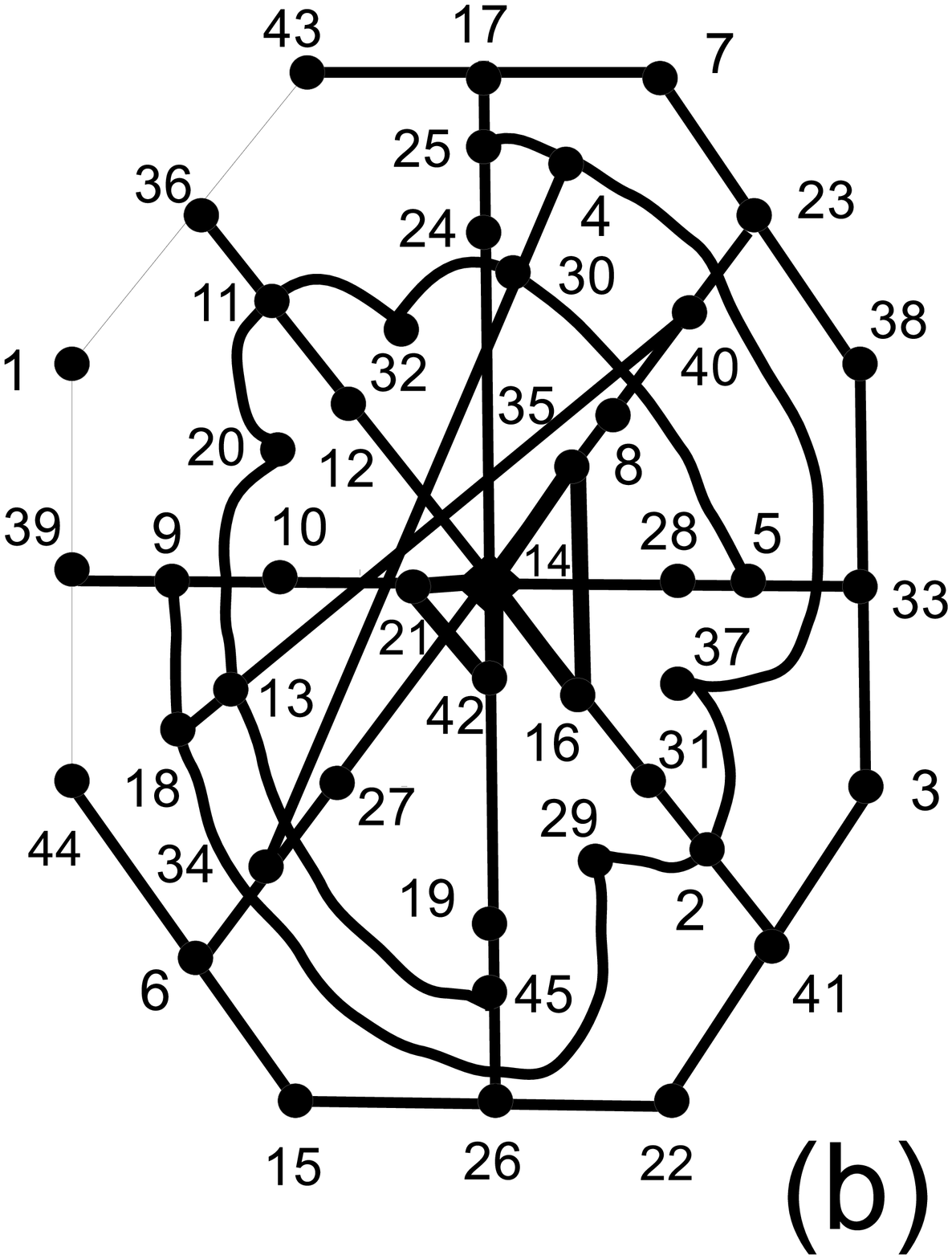}
\caption{The dessin in (a)  arising from the unique subgroup  of index $45$ isomorphic to $A_6$ of the Kleinian group $K^{(5)}$(white points are implicit in the drawing). This dessin stabilizes the thin generalized octagon $GO(2,1)$ shown in (b).  The vertices of the two thicker triangles at the center of the picture (b) form lines $(8,14,16)$ and $(14,21,42)$. They share the point $14$ (the big bullet) and the latter does not lie on either straight line around it. More precisely, there are three lines left along a straight segment going through the point $14$ such as $(6,34,27)$, $(27,35,8)$ and $(8,40,23)$. The list of coset representatives  corresponding to the labeling $1$ to $45$ is
$[e, y, y   x, y   x^{-1}, y   x^2, y   x   y, y   x^{-1}   y, y   x^2   y, (y  x)^2, y   x   y   x^{-1}, y   x^{-1}   y   x, (y   x^{-1})^2, y   x^2   y   x, y  
x^2   y   x^{-1},\newline y   x   y   x^2, y   x   y   x^{-1}   y, y   x^{-1}   y   x^2, y  
x^{-1}   y   x   y, (y   x^2)^2, y   x^2   y   x   y, y   x^2   y   x^{-1}   y, y
x   y   x^2   y, y   x   y   x^{-1}   y   x,\newline y   x   y   x^{-1}   y   x^{-1}, y   x^{-1}
  y   x^2   y, y   x^{-1}   y   x   y   x, y   x^{-1}   y   x   y   x^{-1}, y   x   y
  x^2   y   x, y   x   y   x^2   y   x^{-1}, y   x   y   x^{-1}   y   x   y,\newline y  
x^{-1}   y   x^2   y   x, y   x^{-1}   y   x^2   y   x^{-1}, y   x^{-1}   y   x   y  
x^{-1}   y, y   x   y   x^2   y   x^2, y   x   y   x^2   y   x   y, y   x   y  
x^2   y   x^{-1}   y, \newline y   x   y   x^{-1}   y   x   y   x, y   x^{-1}   y   x^2   y  
x^2, y   x^{-1}   y   x^2   y   x   y, y   x^{-1}   y   x^2   y   x^{-1}   y, y   x^{-1}
  y   x   y   x^{-1}   y   x^{-1}, y   x   y   x^2   y   x   y   x^{-1},\newline  y   x   y  
x^2   y   x^{-1}   y   x, y   x^{-1}   y   x^2   y   x   y   x, y   x^{-1}   y   x^2 
y   x^{-1}   y   x]$. As in previous diagrams, thick lines have non-commuting cosets.
 }
\end{figure}

\subsubsection*{The Ree-Tits octagon}

Thick generalized polygons are stabilized by appropriate dessins d'enfants, as for $GQ(2,2)$, $GQ(2,4)$, $GH(2,2)$ and its dual (that reproduce the commutation structure of two- and three-qubit observables \cite{Planat2015}-\cite{Dessins2014},\cite{PSH}) as well as the Ree-Tits octagon $GO(2,4)$ \cite{Moonshine2015}. But these thick polygons cannot be stabilized from subgroups of Kleinian groups in the $MM$-census: only thin polygons can be recovered.

To illustrate this matter, let us consider the octagon $GO(2,4)$. Take the subgroup $G_1$ of the modular group $\Gamma$ with representation
$$G_1=\left\langle x,y| x^2=y^3=(xy)^{13}=[x,y]^5=[x,yxy]^4=[(xy)^4(xy^{-1})]^6=1\right\rangle$$
and the subgroup $H_1$ of $G_1$ defined as
$$H_1=\mbox{sub}\left\langle G_1|x=y^{-1}(xy)^2xy^{-1}(xy)^3(xy^{-1})^2=1\right\rangle.$$
The index of $H_1$ in $G_1$ is $1755$ and the corresponding permutation group is the Tits group $T$ of order $17,971,200$ \cite{Atlasv3}. The signature of the dessin is $(B,W,F,g)=(1846,1170,270,113)$ with cycles $[2^{1664}1^{82},3^{1170},13^{270}]$, see also \cite[Table 2 and Fig. 5]{Moonshine2015}.

Finally, the stabilizer subgroup of order $2^{10}$ of $T$ is used to recover $GO(2,4)$ that has $1755$ vertices, $2925$ lines/triangles and a collinearity graph of spectrum $[10^1,5^{351},1^{650},-3^{675},-5^{78}]$. The subgroup $H_1$ is found to contain a \lq kernel' which is the Kleinian group $K^{(1)}$ (with the switch from $y$ to $x$ compared to the definition in Sec. 4). The group $K^{(1)}$ has index $4,492,800=2560 \times 1755$ in $G_1$. Schematically
$$G\supset G_1 \supset_{T~ \mbox{from}~\mbox{index}~1755} H_1 \supset_{2560} K^{(1)}.$$

It is known that the thin octagon $GO(2,1)$ is embedded in $GO(2,4)$ in an essentially unique way \cite{DeBruyn2011}. But the supset structure above does not reflect that $GO(2,4)\supset GO(2,1)$ since $GO(2,1)$ follows from the Kleinian group $K^{(5)}$ not $K^{(1)}$.

\subsubsection*{The Cohen-Tits near octagon}

There exists $280$ copies of the thin octagon $GO(2,1)$ within the Cohen-Tits near octagon on $315$ points \cite{CohenTits1985}. This construction is related to that of the Hall-Janko group $J_2$.
Starting with the subgroup $G_2$ of the modular group
$$G_2=\left\langle x,y| x^2=y^3=(xy)^{7}=[x,y]^{12}=[(xy)^2xy^{-1}xy(xy^{-1})^2 (xy)^2 (xy^{-1})^2xyxy^{-1}]^3 =1\right\rangle$$
and the subgroup $H_2$ of $G_2$ defined as
$$H_2=\mbox{sub}\left\langle G_2|y=(xy)^2xy^{-1}xyx=1\right\rangle,$$
the permutation group associated to the coset structure of $H_2$ in $G_2$ is the Hall-Janko group $J_2$ of order $604,800$. One uses the stabilizer subgroup of $H_2$ isomorphic to $PSL(2,7)$ to stabilize the Hall-Janko graph of spectrum $[36^1,6^{36},(-4)^{63}]$. 
The subgroup $H_2$ contains a \lq kernel' in the form of the Kleinian group $K^{(2)}$ defined as
$$K^{(2)}=\left\langle x,y|  y^2=x ^3=[(yx^{-1})^3(y^{-1}x)^3]^3=1\right\rangle$$ so that
$$G\supset G_2 \supset_{J_2~ \mbox{from}~\mbox{index}~100} H_2 \supset_{3024} K^{(2)}.$$

But the supset structure we are concerned with is that used to stabilize the Cohen-Tits near octagon
$$G\supset G_2 \supset_{J_2.2~ \mbox{from}~\mbox{index}~315} H_2' .$$
The permutation representation (isomorphic to $J_2.2$) of the coset structure of $H_2'$ in $G_2$ is available in \cite{Atlasv3} (but not the explicit representation of $G_2$). The stabilizer subgroup of order $192$ of $J_2.2$ is used to recover the Cohen-Tits near octagon on $315$ points with $525$ triangles and spectrum $10^15^{36}3^{90}(-2)^{160}(-5)^{28}$.

\section{Summary}

The coset parametrization of quantum observables (from a subgroup $H$ of the free group $G=\left\langle x,y \right\rangle$ and the dessin d'enfant methodology) allowed us arrive at the result that geometric contextuality fits quantum contextuality \cite{Planat2015}. Here we have shown that non-compact arithmetic Kleinian groups $K=\left\langle f,g \right\rangle$ generated by two elliptic isometries $f$ and $g$ act as \lq contextuality filters'.  Thin generalized polygons $GQ(2,1)$, $GH(2,1)$ and $GO(2,1)$ (and a few extra cases) when stabilized thanks to the relevant finite index subgroups of Kleinian groups $K^{(i)}$ in the Maclachlan-Martin census \cite{Maclachlan2001} have been shown to be maximally contextual. All hyperbolic orbifolds $\mathbb{H}^3/K^{(i)}$ are candidates for a re-examination of geometric contextuality with the further remark that the $K^{(i)}$'s are commensurable with Bianchi groups $PSL(2,O_d)$, $d \in \{1,3\}$, whose elements are in the ring of integers of $\mathbb{Q}(\sqrt{-d})$.

\end{document}